\documentclass[preprintnumbers,amsmath,11pt,amssymb,floatfix,superscriptaddress,nofootinbib]{article}

\def\mytitle#1{\setcounter{equation}{0}
\setcounter{footnote}{0}
\begin{flushleft}\Large\textbf{#1}\end{flushleft}
\vspace{0.25cm}}
\def\myname#1{\leftline{{\large #1}}\vspace{-0.13cm}}
\def\myplace#1#2{\small\begin{flushleft}\textit{#1}\\
\texttt{#2}\end{flushleft}}

\usepackage{graphicx}
\begin{document}

\mytitle{Dynamics of Modified Chaplygin Gas in Brane World
Scenario: Phase Plane Analysis}

\vskip0.2cm \myname{ Prabir
Rudra~\footnote{prudra.math@gmail.com}$*$} \vskip0.2cm \myname{
Ujjal Debnath\footnote{ujjal@iucaa.ernet.in,
ujjaldebnath@yahoo.com}~$*$} \vskip0.2cm \myname{ Ritabrata
Biswas\footnote{biswas.ritabrata@gmail.com}~$\dag$}

\myplace{$*$Department of Mathematics, Bengal Engineering and
Science University, Shibpur, Howrah-711 103, India.}
{}
\myplace{$\dag$ Department of Mathematics, Jadavpur University,
Kolkata-700 032, India.} {}

\begin{abstract}
In this work we investigate the background dynamics when dark
energy is coupled to dark matter with a suitable interaction in
the universe described by brane cosmology. Here DGP and the RSII
brane models have been considered separately. Dark energy in the
form of modified Chaplygin gas is considered. A suitable
interaction between dark energy and dark matter is considered in
order to at least alleviate (if not solve) the cosmic coincidence
problem. The dynamical system of equations is solved numerically
and a stable scaling solution is obtained. A significant attempt
towards the solution of the cosmic coincidence problem is taken.
The statefinder parameters are also calculated to classify the
dark energy models. Graphs and phase diagrams are drawn to study
the variations of these parameters. It is also seen that the
background dynamics of modified Chaplygin gas is completely
consistent with the notion of an accelerated expansion in the late
universe. Finally, it has been shown that the universe in both the
models follows the power law form of expansion around the critical
point, which is consistent with the known results.
\end{abstract}

\section{Introduction}

The idea of the static universe once conceived by Albert Einstein
faced a big threat when at the turn of the last century
observations of Ia Supernova confirmed that our universe is
suffering from an accelerated expansion (Perlmutter, S. et al.,
1999; Spergel, D. N. et al.2003). Now belief and observations are
always to be supported by some model. In the quest of finding a
suitable model for universe, Cosmologists started to investigate
the root cause that is triggering this expansion. Fundamentally,
we were to modify Einstein's equation either by modifying the left
hand side, i.e., modifying the idea of Einstein gravity or to
modify the right hand side which immediately speculate the nature
of the matter inside the universe. If our Universe is filled up by
some invisible fluid causing a negative pressure then it violates
the strong energy condition i.e. $\rho+3p<0$. Because of its
invisible nature this energy component was aptly termed as dark
energy (DE) (Riess, A. G. et al., 2004).

With the introduction of DE,  the search began for different
candidates that can effectively play the role of DE.  DE
represented by a scalar field \footnote{ in the presence of a
scalar field the transition from a universe filled with matter to
an exponentially expanding universe is justified } is often called
quintessence. Not only scalar field but also there are other Dark
fluid models like Chaplygin gas which plays the role of DE very
efficiently.  The earliest form of this was known as pure
Chaplygin gas (Kamenshchik, A. et al.,2001; Gorini, V. et al.
2004). Extensive research saw pure Chaplygin gas first modify into
generalized Chaplygin gas (Gorini, V. et al., 2006; Alam, U. et
al. 2003; Bento, M. C. et al. 2002;Carturan, D. and Finelli, F.
2003; Barreiro, T. and  Sen, A.A. 2004) and later into modified
Chaplygin gas (MCG) obeying the equation of state (Benaoum, H. B.,
2002; Debnath, U., 2004)

\begin{equation}\label{2}
p=A\rho-\frac{B}{\rho^\alpha},
\end{equation}
where $p$ and $\rho$ are respectively the pressure and energy
density and $0\leq\alpha\leq1$, $A$ and $B$ are positive
constants. Currently, we live in a special epoch where the
densities of DE and DM are comparable. Although they have evolved
independently from different mass scales. This is known as the
famous cosmic coincidence problem. till date several attempts have
been made to find a solution to this problem (del Campo, S. et al
2009; Leon, G. and Saridakis, E.N. 2010; Jimenez, J.B. and Maroto,
A.L.2008; Berger, M.S. and Shojae, H. 2006; Zhang, X. 2005;
Griest, K. 2002; Jamil, M. and Rehman, F. 2009; Jamil, M. et al.
2010a; Jamil, M. and Saridakis, E. N. 2010b; Jamil, M. and Farooq,
M. U., 2010c; Jamil, M. et al. 2010d). A suitable interaction
between DE and DM is required if we wish to find an effective
solution to this problem. It is obvious that there has been a
transition from a matter dominated universe to dark energy
dominated universe, by exchange of energy at an appropriate rate.
Now the expansion history of the universe as determined by the
supernovae and CMB data (Jamil, M.et al. 2010a; Jamil, M. and
Saridakis, E. N. 2010b) bounds us to fix the decay rate such that
it is proportional to the present day Hubble parameter. Keeping
the fact in mind cosmologists all over the world have studied and
proposed a variety of interacting DE models (Setare,  M. R.
2006,2007c; Hu, B. and Ling, Y. 2006; Wu, P. and Yu, H. 2007;
Jamil, M. 2010e, 2011; Dalal, N. 2001).

Here we shall use MCG as the DE and the equation of state for MCG
is given by (\ref{2}) with $\rho=\rho_{mcg}$
Recent developments regarding MCG can be found in (Lu, J. et al.
2008; Dao-Jun, L., Xin-Zhou,  L. 2005; Jing, H. et. al. 2008;
Debnath, U. 2007). This special form also appears to be consistent
with the WMAP 5-year data and henceforth support the unified model
with DE and DM based on generalized CG (Makler, M. et al. 2003;
Setare, M.R. 2007a, 2007b, 2009; Barreiro, T. et al. 2008). Recent
supernovae data also favours the two-fluid cosmological model with
CG and matter (Panotopoulos, G. 2008).

As we have stated earlier, modifying the right hand side of
Einstein's equation was not the only way to explain the increase
in the rate of the expansion, rather to modify the gravity part of
the left hand side is also to demonstrate the present day
universe. In this branch Brane-gravity was introduced and brane
cosmology was developed. A review on brane-gravity and its various
applications with special attention to cosmology is available in
(Rubakov, V. A. 2001; Maartens, R. 2004; Brax. P. et al. 2004; Csa
ki, C. 2004). In this work we consider the two most popular brane
models, namely DGP and RS II branes. Our main aim of this work is
to examine the nature of the different physical parameters for the
universe around the stable critical points in two brane world
models in presence of MCG. Impact of any future singularity caused
by the DE in brane world models will be studied.

This paper is organized as follows: Section 2 comprises of the
dynamical system analysis in DGP brane model. Section 3 deals with
the dynamical system analysis in RS II brane model, followed by a
general discussion in section 4.

\section{Dynamical system analysis in DGP brane world model}

A simple and effective model of brane-gravity is the
Dvali-Gabadadze-Porrati (DGP) braneworld model (Dvali, G. R. et
al. 2000; Deffayet, D. 2001; Deffayet, D. et al. 2002) which
models our 4-dimensional world as a FRW brane embedded in a
5-dimensional Minkowski bulk. It explains the origin of DE as the
gravity on the brane leaking to the bulk at large scale. On the
4-dimensional brane the action of gravity is proportional to
$M_p^2$ whereas in the bulk it is proportional to the
corresponding quantity in 5-dimensions. The model is then
characterized by a cross over length scale $
r_c=\frac{M_p^2}{2M_5^2} $ such that gravity is 4-dimensional
theory at scales $a<<r_c$ where matter behaves as pressureless
dust, but gravity leaks out into the bulk at scales $a>>r_c$ and
matter approaches the behaviour of a cosmological constant.
Moreover it has been shown that the standard Friedmann cosmology
can be firmly embedded in DGP brane.

It may be noted that in literature, standard DGP model has been
generalized to (i) LDGP model by adding a cosmological constant
(Lue,A. and Starkman,G. D. 2004), (ii) QDGP model by adding a quintessence perfect
fluid (Chimento,L. P. et al. 2006), (iii) CDGP model by Chaplygin gas
(Bouhmadi-Lopez, M. and Lazkoz, R. 2007) and (iv) SDGP by a scalar field
(Zhang, H. and Zhu, Z. H. 2007). In (Wu, X. et al. 2008) the DGP model has been analysed by
adding Holographic DE (HDE).
\subsection{Basic equations in DGP brane model}\label{sec2}

While flat, homogeneous and isotropic brane is being considered, the Friedmann equation
 in DGP brane model(Dvali, G. R. et al, 2000; Deffayet, D., 2001; Deffayet, D. et al,
2002) is modified to the equation
\begin{equation}\label{3}
H^2=\left(\sqrt{\frac{\rho}{3}+\frac{1}{4r_{c}^{2}}}+\epsilon \frac{1}{2r_c}\right)^2,
\end{equation}
where $H=\frac{\dot a}{a}$ is the Hubble parameter, $\rho$ is the
total cosmic fluid energy density and $r_c=\frac{M_p^2}{2M_5^2}$
is the cross-over scale which determines the transition from 4D to
5D behaviour and $\epsilon=\pm 1 $ (choosing $M_{p}^{2}=8\pi
G=1$). For $\epsilon=+1$, we have standard DGP$(+)$ model which is
self accelerating model without any form of DE, and effective $w$
is always non-phantom. However for $\epsilon=-1$, we have DGP$(-)$
model which does not self accelerate but requires DE on the brane.
It experiences 5D gravitational modifications to its dynamics
which effectively screen DE. Brane world scenario is actually a
modified gravity theory. If we write the Einstein equation for
brane world in terms of Einstein gravity then the extra term can
be treated as the effective DE But that is not the physical DE.
Moreover this DE is applicable only in Einstein gravity. But here
we will consider the physical DE in brane world. So we have
introduced the MCG type fluid in brane. There are few many
candidates of DE models, such as scalar field (Nojiri, S.and
Oditsov, S.D. 2004), phantom (Nojiri, S. et al 2005), k-essence
(Bamba, K.et al 2011), tachyonic field ((Nojiri, S.and Oditsov,
S.D. 2003), etc. MCG being a modified form of DE with a unique
equation of state for DE and DM. Also MCG is the observational
viable model (Thakur, P. et al 2009, Dev, A. et al 2003). It
interpolates between radiation in early stage and $\Lambda$CDM in
late stage.

As in the present problem the interaction between DE and pressureless
dark matter has been taken into account for interacting DE and DM the energy balance equation will be
\begin{equation}\label{4}
\dot{\rho}_{mcg}+3H\left(1+w_{mcg}\right)\rho_{mcg}=-Q,~~~for ~MCG~ and ~
\end{equation}
\begin{equation}\label{5}
\dot{\rho}_m+3H\rho_m=Q, ~for~ the~ DM~ interacting ~with~ MCG.
\end{equation}
where $Q=3bH\rho$ is the interaction term, $b$ is the coupling
parameter (or transfer strength) and $\rho=\rho_{mcg}+\rho_m$ is
the total cosmic energy density which satisfies the energy
conservation equation $\dot{\rho}+3H\left(\rho+p\right)=0$ (Guo,
Z.-K. and Zhang, Y.-Z. 2005; delCampo, S. et. al. 2008).

As we have lack of information about the fact how do DE and DM interact so we are not able to esteemate the interaction term from the first principles. However, the negativity of $Q$ immediately implies the possibility of having negative DE in the early universe which is overruled by to the necessity of the second law of thermodynamics to be held(Alcaniz, J. S. and Lima, J. A. S. 2005). Hence $Q$ must be
positive and small. From the observational data of 182 Gold type
Ia supernova samples, CMB data from the three year WMAP survey and
the baryonic acoustic oscillations from the Sloan Digital Sky
Survey, it is estimated that the coupling parameter between DM and
DE must be a small positive value (of the order of unity), which
satisfies the requirement for solving the cosmic coincidence
problem and the second law of thermodynamics (Feng, C. et. al.
2008). Due to the underlying interaction, the beginning of the
accelerated expansion is shifted to higher redshifts. Consequently
using the Friedmann equation (\ref{3}) and the conservation
equation, we obtain the modified Raychaudhuri equation
\begin{equation}\label{6}
\left(2H-\frac{\epsilon}{r_c}\right)\dot{H}=-H\left(\rho+p\right),
\end{equation}
\subsection{Dynamical system analysis}
In this subsection we plan to analyse the dynamical system. For that firstly the conversion of the physical parameters into some dimensionless form, given by
\begin{equation}\label{7}
x=\ln a, ~~ u=\frac{\rho_{mcg}}{3H^2}, ~~ v=\frac{\rho_m}{3H^2}
\end{equation}
where the present value of the scale factor $a_0=1$ is assumed.
Inclusion of (\ref{2}) into (\ref{7}) gives us the parameter gradients as
\begin{equation}\label{8}
\frac{du}{dx}=-3b\left(u+v\right)-3u\left(1+w_{mcg}^{(DGP)}\right)+
\frac{3u\left\{\epsilon-\sqrt{1+(\epsilon^2-1)(u+v)}\right\}\left\{u(1+w_{mcg}^{(DGP)})+v\right\}}{\epsilon\left(u+v\right)-\sqrt{1+(\epsilon^2-1)(u+v)}}
\end{equation}
and
\begin{equation}\label{9}
\frac{dv}{dx}=3b(u+v)-3v+\frac{3v\left\{\epsilon-\sqrt{1+(\epsilon^2-1)(u+v)}\right\}\left\{u(1+w_{mcg}^{(DGP)})+v\right\}}{\epsilon\left(u+v\right)-\sqrt{1+(\epsilon^2-1)(u+v)}}.
\end{equation}
Where, $w_{mcg}^{(DGP)}$ is the EoS parameter for MCG determined
as
\begin{equation}\label{10}
w_{mcg}^{(DGP)}=\frac{p_{mcg}}{\rho_{mcg}}=A-B\left[\frac{ 2^{2}
r_c^{2} (1-u-v)^{2}}{3 u
\left(\epsilon-\sqrt{1+(\epsilon^2-1)(u+v)}\right)^{2}}\right]^{(\alpha+1)}.
\end{equation}

\subsubsection{Critical point}

For mathematical simplicity, we consider $\alpha=1$ only. The
critical points of the above system are obtained by putting
$\frac{du}{dx}=0=\frac{dv}{dx}$. Now this system of equation does
not yield an explicit solution. So we have to put some special
values to the constants which will yield a non zero positive
solution of the system of equations. Considering $A=0$ and
$\epsilon=1$ we obtain the following critical value to the system.

\begin{equation}\label{11}
u_{c}=1-b-\frac{3D}{2Br^4}-\frac{\sqrt{3}}{2}\sqrt{\frac{3\left(1-3b+3b^2-b^3\right)}{Br^4}-\frac{4\left(1+4b-6b^2+4b^3-b^4\right)}{D}}
\end{equation}
\begin{equation}\label{12}
v_{c}=\frac{1}{Br^4}\left(\frac{9-18b+9b^2}{u_{c}}+\frac{3}{2}D\right)+3+b+\frac{\sqrt{3N}}{2}-\frac{1+3b-3b^2+b^3}{u_{c}^3}+\frac{4\left(1-2b+b^2\right)}{u_{c}^2}-\frac{6\left(1+b\right)}{u_{c}}
\end{equation}

where $$D=\sqrt{B\left(1-3b+3b^2-b^3\right)}~r^2$$ and

$$N=\left\{\frac{2}{\sqrt{3}}\left(u_{c}+b-1+\frac{3\sqrt{1-3b+3b^2-b^3}}{2\sqrt{B}r^2}\right)\right\}^2$$

The critical point correspond to the era dominated by DM and MCG
type DE. For the critical point $(u_{c},v_{c})$, the equation of
state parameter (\ref{10}) of the interacting DE takes the form
\begin{equation}\label{13}
w_{mcg}^{(DGP)}=A-\frac{B2^{2(\alpha+1)}
r_c^{2(\alpha+1)}(1-u_{c}-v_{c})^{2(\alpha+1)}}{3^{\alpha+1}u_{c}^{\alpha+1}
(\epsilon-\sqrt{1+(\epsilon^2-1)(u_{c}+v_{c})})^{2(\alpha+1)}}.
\end{equation}

\subsubsection{Stability around critical point}

Now we check the stability of the dynamical system (eqs. (\ref{8})
and (\ref{9})) about the critical point. In order to do this, we
linearize the governing equations about the critical point i.e.,
\begin{equation}\label{14}
u=u_c+\delta u ~~  and ~~  v=v_c+\delta v,
\end{equation}
Now if we assume $f=\frac{du}{dx}$ and $g=\frac{dv}{dx}$, then we
may obtain
\begin{equation}\label{15}
\delta\left(\frac{du}{dx}\right)=\left[\partial_{u}
f\right]_{c}\delta u+\left[\partial_{v} f\right]_{c}\delta v
\end{equation}
and
\begin{equation}\label{16}
\delta\left(\frac{dv}{dx}\right)=\left[\partial_{u}
g\right]_{c}\delta u+\left[\partial_{v} g\right]_{c}\delta v
\end{equation}

where

$$\partial_{u} {f}= \left[-3b-3Bu\left(\frac{-32\theta^4\left(\epsilon^2-1\right)r_c^4}{9u^2K\left(\epsilon-K\right)^5}+\frac{64\theta^3r_c^4}{9u^2\left(\epsilon-K\right)^4}
+\frac{32\theta^4r_c^4}{9u^3\left(\epsilon-K\right)^4}\right)-3\left(1+A-\frac{16B\theta^4r_c^4}{9u^2\left(\epsilon-K\right)^4}\right)\right.$$
$$+\frac{1}{\epsilon\left(u+v\right)-3uK\left(\epsilon-K\right)}\left[1+A-\frac{16B\theta^4r_c^4}{9u^2\left(\epsilon-K\right)^4} +u\left(-\frac{32B\theta^4\left(\epsilon^2-1\right)r_c^4}{9u^2K\left(\epsilon-K\right)^5}+\frac{64B\theta^3r_c^4}{9u^2\left(\epsilon-K\right)^4}+\frac{32B\theta^4r_c^4}{9u^3\left(\epsilon-K\right)^4}\right)\right]$$
$$-\frac{3u\left(\epsilon-\frac{\epsilon^2-1}{2K}\right)\left(\epsilon-K\right)\left(v+u\left(1+A-\frac{16B\theta^4r_c^4}{9u^2\left(\epsilon-K\right)^4}\right)\right)}{\left(\left(u+v\right)\epsilon-K\right)^2}-\frac{3u\left(\epsilon^2-1\right)\left(v+u\left(1+A-\frac{16B\theta^4r_c^4}{9u^2\left(\epsilon-K\right)^4}\right)\right)}{2K\left(\left(u+v\right)\epsilon-K\right)}$$
\begin{equation}\label{17}
~~~~~~~~~~~~~~~~~~~~~~~~~~~~~~~~~~~~~~~~~~~~~~~~~~~~~~~~~~~~~~~\left.+\frac{3\left(\epsilon-K\right)\left(v+u\left(1+A-\frac{16B\theta^4r_c^4}{9u^2\left(\epsilon-K\right)^4}\right)\right)}{\left(u+v\right)\epsilon-K}\right]
\end{equation}

$$\partial_{v}{f}=\left[-3b-3u\Big(-\frac{32B\theta^4\left(\epsilon^2-1\right)r_c^4}{9u^2K\left(\epsilon-K\right)^5}+\frac{64B\theta^3r_c^4}{9u^2\left(\epsilon-K\right)^4}\right)+\frac{3u\left(\epsilon-K\right)\left(1+u\left(-\frac{32B\theta^4\left(\epsilon^2-1\right)r_c^4}{9u^2K\left(\epsilon-K\right)^5}+\frac{64B\theta^3r_c^4}{9u^2\left(\epsilon-K\right)^4}\right)\right)}{\left(u+v\right)\epsilon-K}$$

\begin{equation}\label{18}
-\frac{3u\left(\epsilon-\frac{\epsilon^2-1}{2K}\right)\left(\epsilon-K\right)\left(v+u\left(1+A-\frac{16B\theta^4r_c^4}{9u^2\left(\epsilon-K\right)^4}\right)\right)}{\left(\left(u+v\right)\epsilon-K\right)^2}\left.-\frac{3u\left(\epsilon^2-1\right)\left(v+u\left(1+A-\frac{16B\theta^4r_c^4}{9u^2\left(\epsilon-K\right)^4}\right)\right)}{2K\left(\left(u+v\right)\epsilon-K\right)}\right]\end{equation}

$$\partial_{u}{g}=
\left[3b+\frac{1}{\epsilon\left(u+v\right)-3vK\left(\epsilon-K\right)}\left[1+A-\frac{16B\theta^4r_c^4}{9u^2\left(\epsilon-K\right)^4}+u\left(-\frac{32B\theta^4\left(\epsilon^2-1\right)r_c^4}{9u^2K\left(\epsilon-K\right)^5}+\frac{64uB\theta^3r_c^4+32B\theta^4r_c^4}{9u^3\left(\epsilon-K\right)^4}\right)\right]\right.$$

\begin{equation}\label{19}
\left.-\frac{3v\left(\epsilon-\frac{\epsilon^2-1}{2K}\right)\left(\epsilon-K\right)\left(v+u\left(1+A-\frac{16B\theta^4r_c^4}{9u^2\left(\epsilon-K\right)^4}\right)\right)}{\left(\left(u+v\right)\epsilon-K\right)^2}\right.\left.-\frac{3v\left(\epsilon^2-1\right)\left(v+u\left(1+A-\frac{16B\theta^4r_c^4}{9u^2\left(\epsilon-K\right)^4}\right)\right)}{2K\left(\left(u+v\right)\epsilon-K\right)}\right]
\end{equation}

$$\partial_{v}{g}=
\left[-3+3b+\frac{3v\left(\epsilon-K\right)\left[1+u\left(-\frac{32B\theta^4\left(\epsilon^2-1\right)r_c^4}{9u^2K\left(\epsilon-K\right)^5}+\frac{64B\theta^3r_c^4}{9u^2\left(\epsilon-K\right)^4}\right)\right]}{\left(u+v\right)\epsilon-K}+\frac{3\left(\epsilon-K\right)\left(v+u\left(1+A-\frac{16B\theta^4r_c^4}{9u^2\left(\epsilon-K\right)^4}\right)\right)}{\left(u+v\right)\epsilon-K}\right.$$

\begin{equation}\label{20}\left.-\frac{3v\left(\epsilon-K\right)}{\left(\left(u+v\right)\epsilon-K\right)^{2}}\left(\epsilon-\frac{\left(\epsilon^2-1\right)}{2K}\right)\left(v+u\left(1+A-\frac{16B\theta^4r_c^4}{9u^2\left(\epsilon-K\right)^4}\right)\right)-\frac{3v\left(\epsilon^2-1\right)\left(v+u\left(1+A-\frac{16B\theta^4r_c^4}{9u^2\left(\epsilon-K\right)^4}\right)\right)}{2K\left(\left(u+v\right)\epsilon-K\right)}\right]\end{equation}

 where $K=\sqrt{1+\left(u+v\right)\left(\epsilon^2-1\right)}$,~~~~~ and
 ~~~~$\theta=\left(1-u-v\right)$. The subscript $c$ refers to quantities evaluated at the critical
point of the dynamical system.

\begin{figure}

\includegraphics[height=2in]{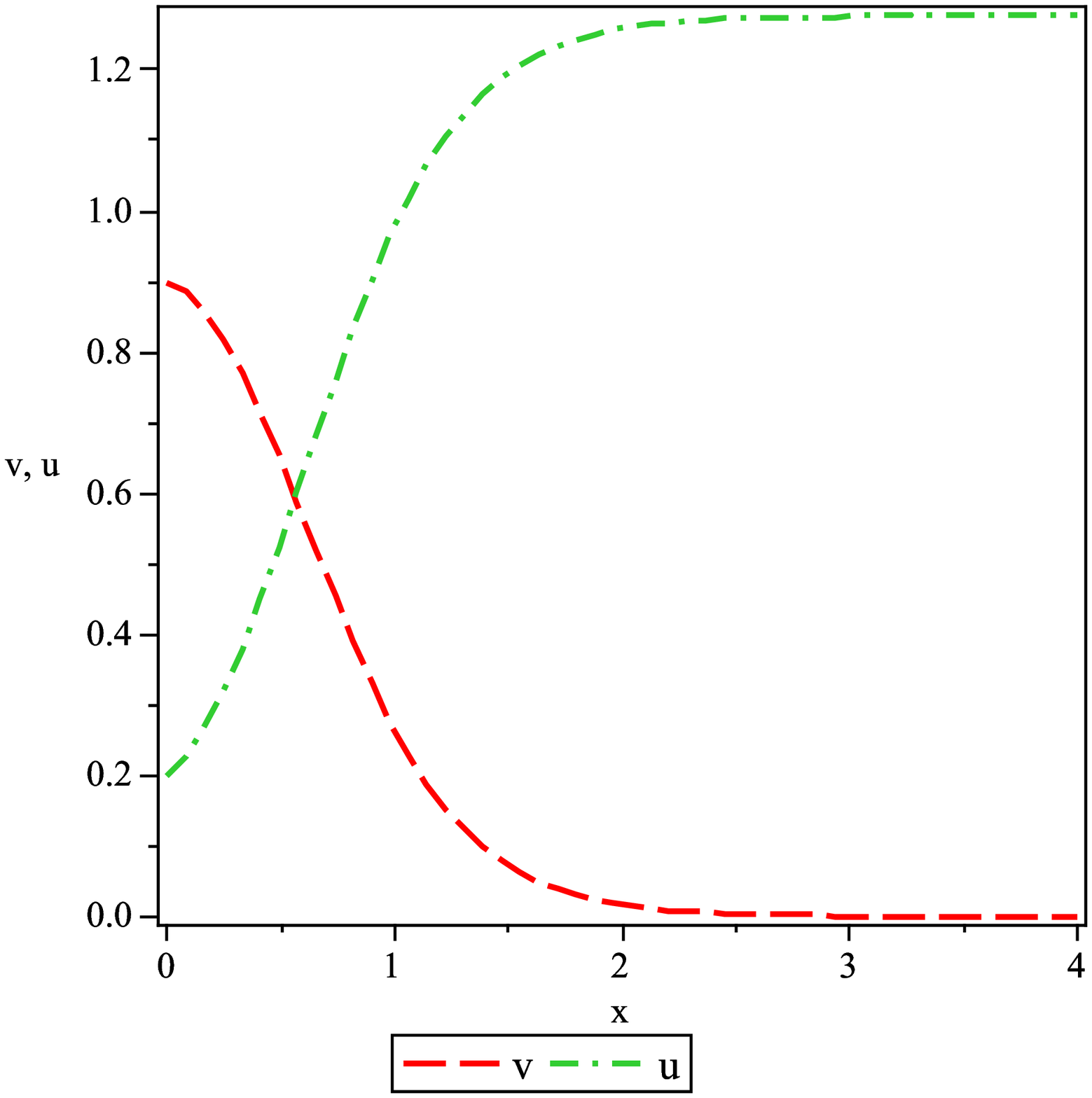}~~~~~~~~\includegraphics[height=2in]{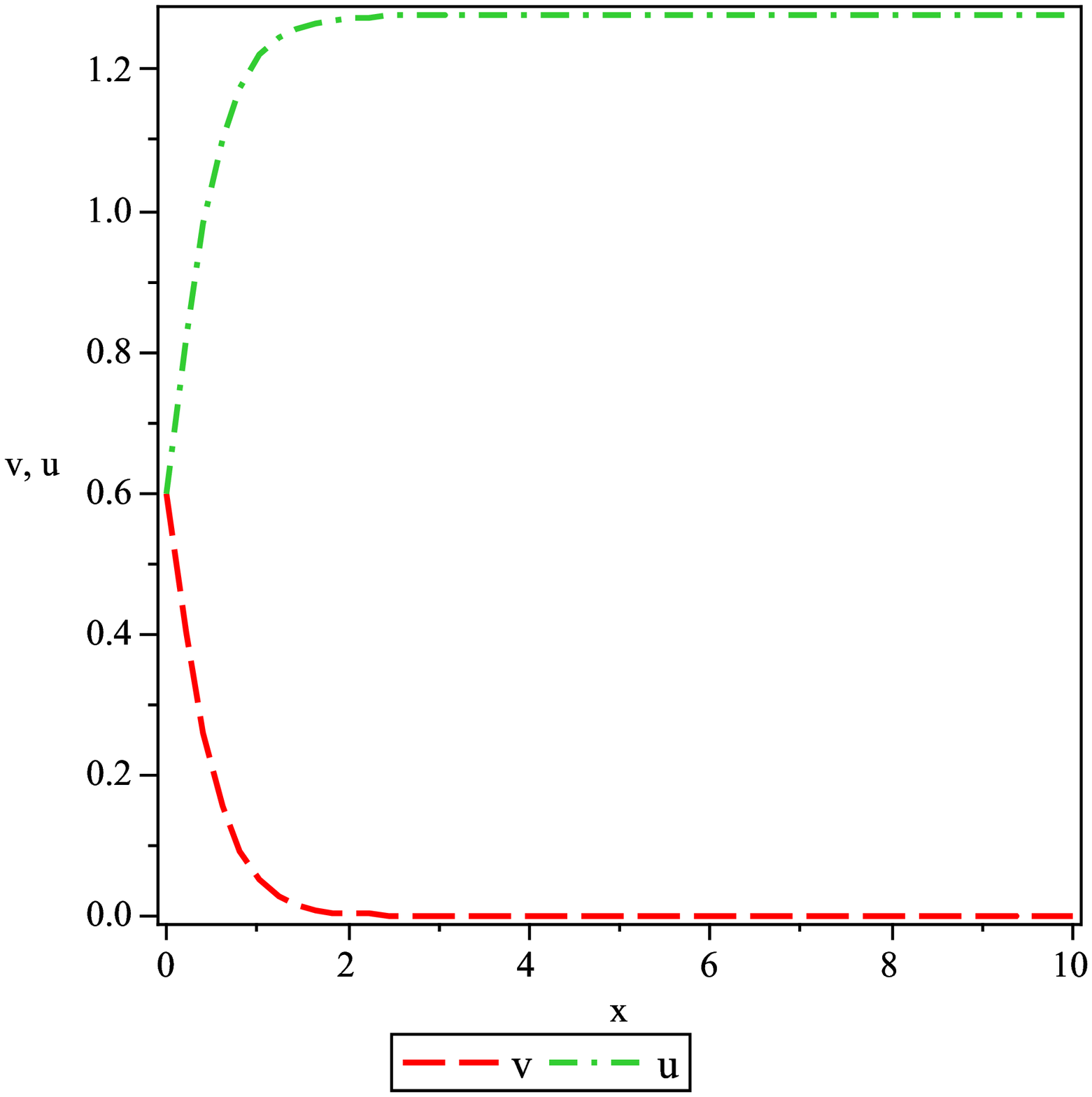}~~~~~~~~\includegraphics[height=2in]{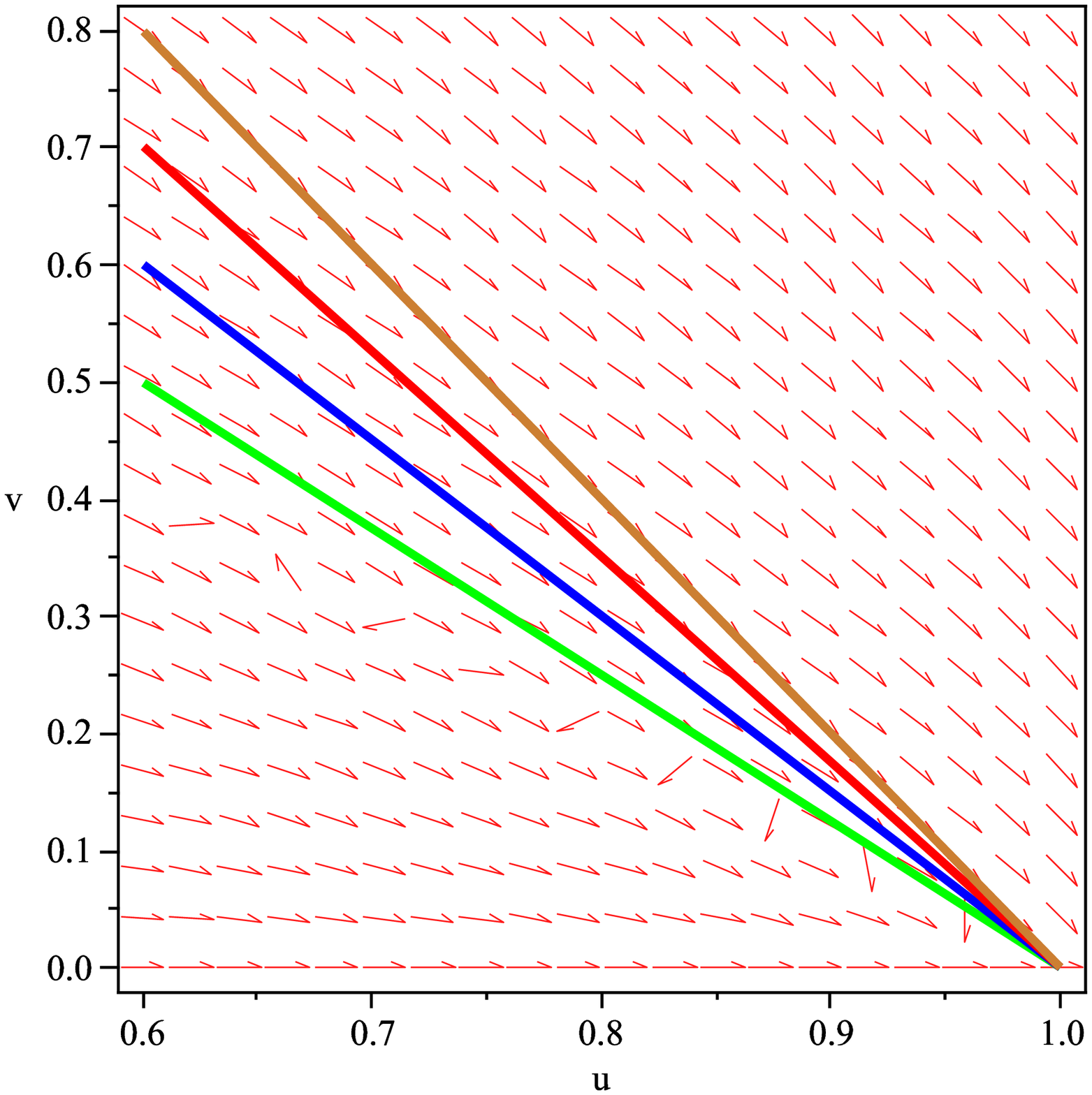}~\\
\vspace{1mm}
~~~~~~~~~~~~~~Fig. 1~~~~~~~~~~~~~~~~~~~~~~~~~~~~~~~~~~~~~~~~Fig. 2~~~~~~~~~~~~~~~~~~~~~~~~~~~~~~~~~~~~~~~~~~Fig. 3~~~\\
\vspace{2mm}

Fig 1 : The dimensionless density parameters are plotted against
e-folding time. The initial condition is $v(0)=0.9, u(0)=0.2$.
Other parameters are fixed at $\alpha=1, b=0, A=1, B=0.5$ and
$r_c=10$.\\
Fig 2 : The dimensionless density parameters are plotted against
e-folding time.The initial condition is $v(0)=0.6, u(0)=0.6$. The
other parameters are fixed at $\alpha=1, b=0, A=1, B=0.5$ and
$r_c=10$.\\Fig 3 :  The phase diagram of the parameters depicting
an attractor solution. The initial conditions chosen are
$v(0)=0.5, u(0)=0.6$ (green); $v(0)=0.6, u(0)=0.6$ (blue);
$v(0)=0.7, u(0)=0.6$ (red); $v(0)=0.8, u(0)=0.6$ (brown). Other
parameters are
fixed at $\alpha=1, b=0, A=0.3, B=1$ and $r_c=10000$.\\\\\\\\

\includegraphics[height=1.5in]{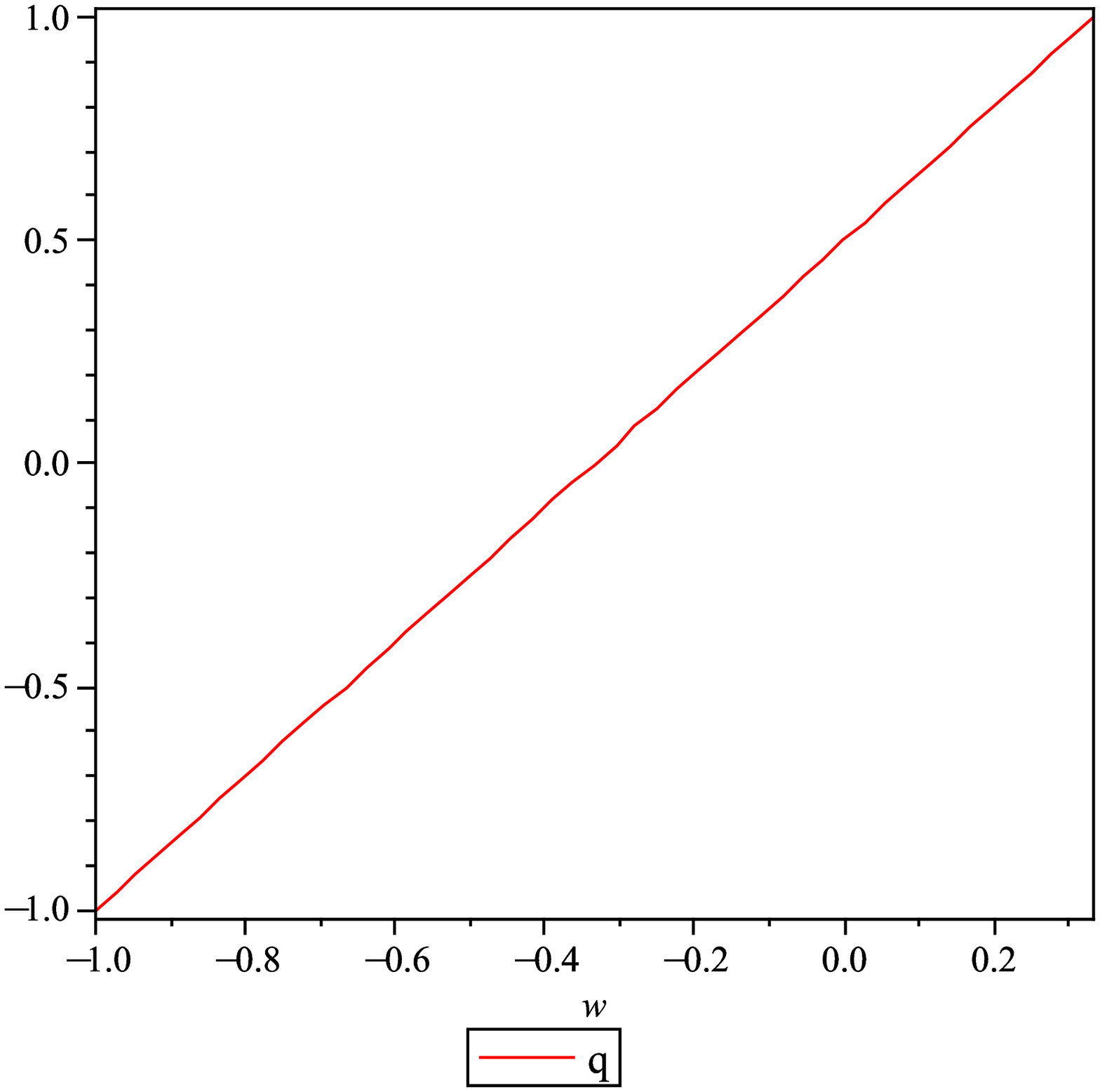}~~~~~\includegraphics[height=1.5in]{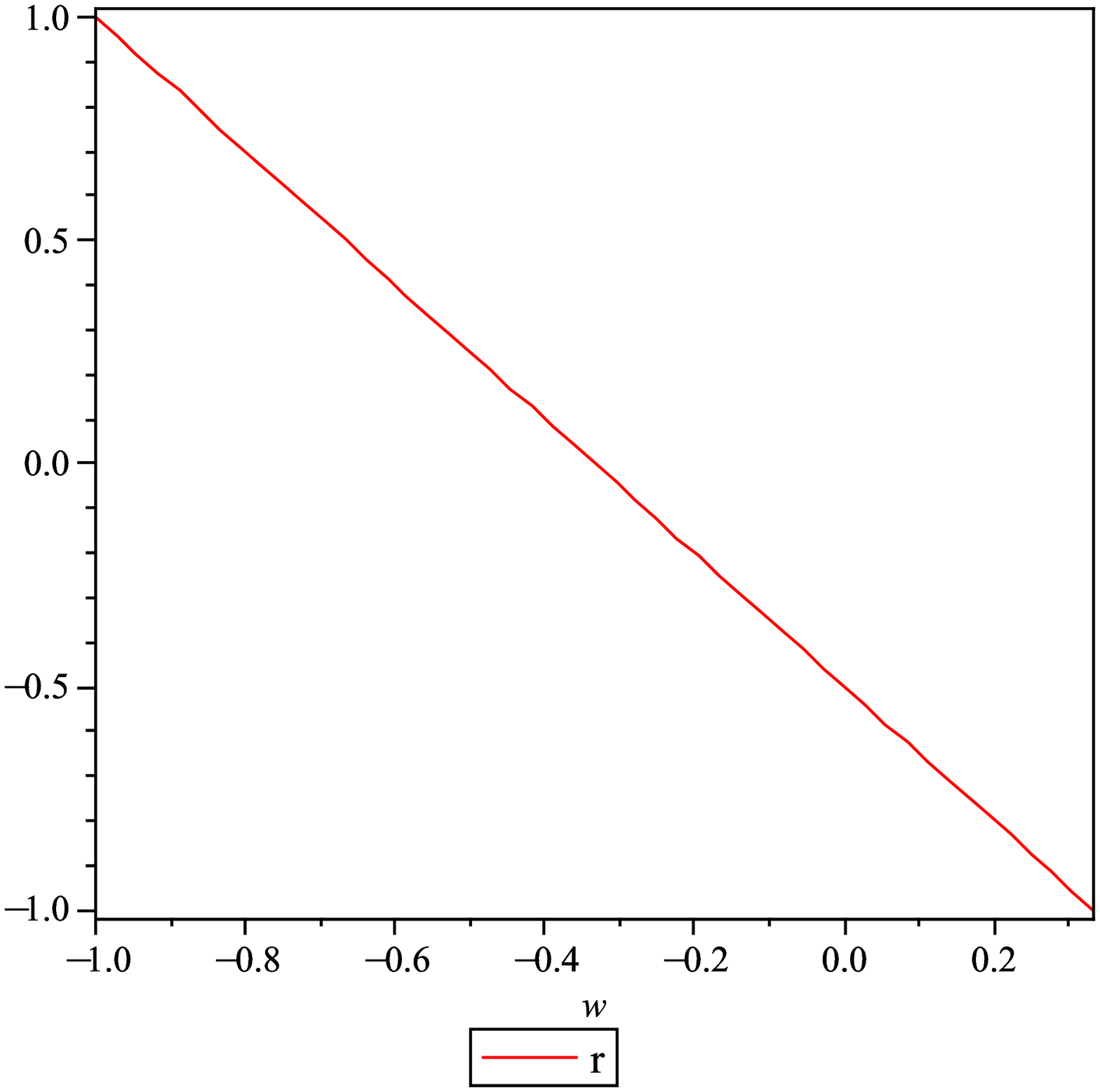}
~~~~~\includegraphics[height=1.5in]{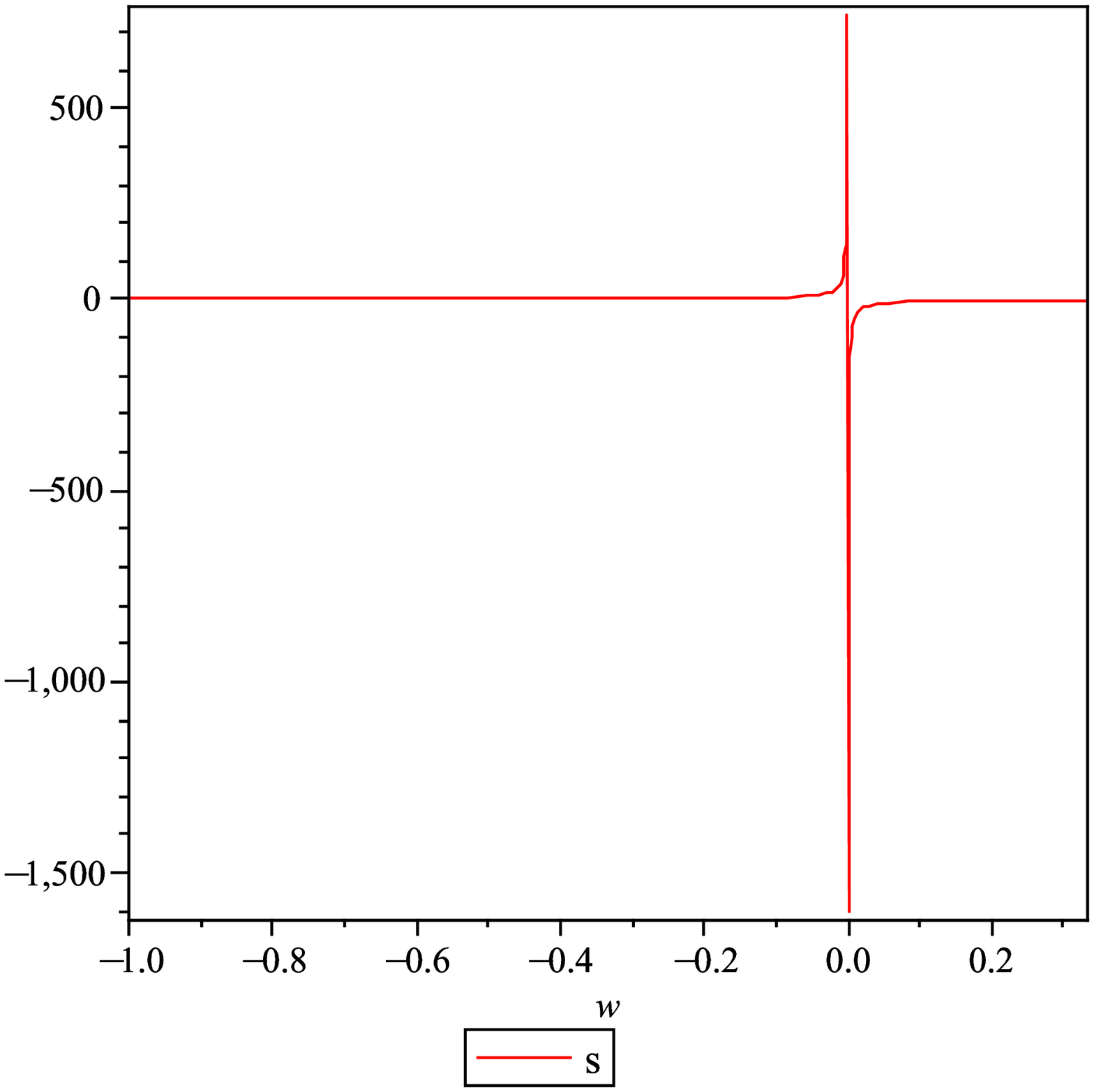}~~~~~\includegraphics[height=1.5in]{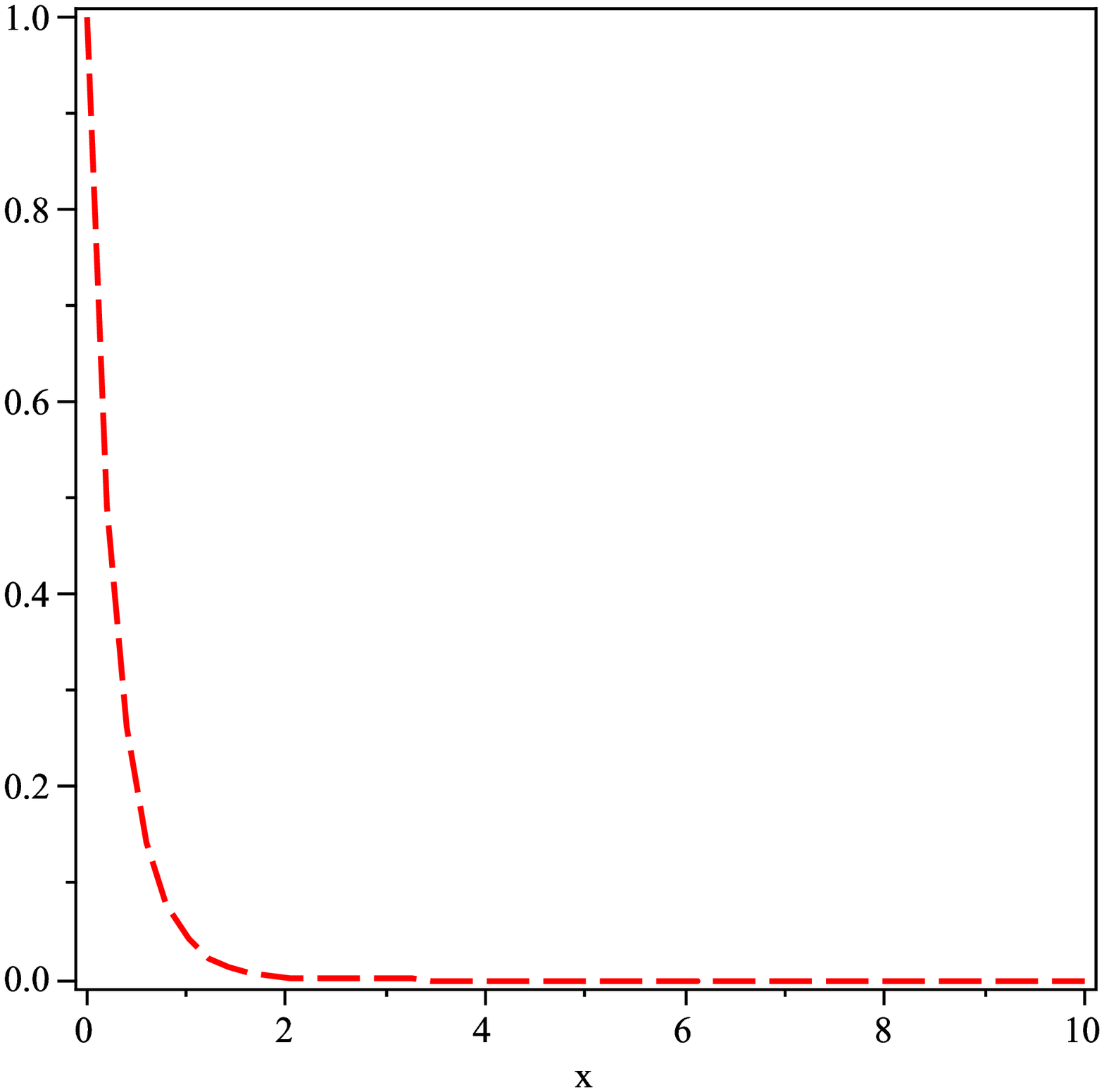}~\\
\vspace{1mm}
~~~~~~~~~~Fig. 4~~~~~~~~~~~~~~~~~~~~~~~~~~~Fig. 5~~~~~~~~~~~~~~~~~~~~~~~~~~~Fig. 6~~~~~~~~~~~~~~~~~~~~~~~~~~~Fig. 7\\
\vspace{2mm}

Fig 4 :   The deceleration parameter is plotted against the state
parameter. Other parameters are fixed at $B=1, b=0, r_c=10000,
\epsilon=-1$.
\\Fig 5 :  The statefinder parameter $r$ is plotted
against the state parameter. The other parameters are fixed at
$B=1, b=0, r_c=10000, \epsilon= -1$.
\\ Fig 6 : The statefinder parameter $s$ is plotted
against the state parameter. The other parameters are fixed at
$B=1, b=0, r_c=10000, \epsilon=-1$.
\\Fig 7 : The ratio of density parameters is shown
against e-folding time. The initial conditions chosen are
v(0)=0.6, u(0)=0.6. The other parameters are fixed at $\alpha=1,
A=1, b=0, B=0.5, r_c=10$.
\end{figure}

\subsubsection{Nature of cosmological parameters}

We calculate the deceleration parameter ~$q=-1-(\dot H/H^2)$, in
this model as,
\begin{equation}\label{21}
q^{(DGP)}=-1+\frac{3}{2}\left\{\frac{12r_c^2\left(1+w_{mcg}^{(DGP)}\frac{\rho_{mcg}}{\rho}\right)}{\sqrt{4r_c^2+3\sigma}\left(\sqrt{4r_c^2+3\sigma}+\frac{\epsilon}{\sqrt{\sigma}}\right)}\right\},
\end{equation}
where $\sigma=\frac{1}{\rho}$~. The above deceleration parameter
can be written in terms of dimensionless density parameter
$\Omega_{mcg}=\frac{\rho_{mcg}}{\rho}$ as
\begin{equation}\label{22}
q^{(DGP)}=-1+\frac{3}{2}\left\{\frac{12r_c^2\left(1+w_{mcg}^{(DGP)}\Omega_{mcg}\right)}{\sqrt{4r_c^2+3\sigma}\left(\sqrt{4r_c^2+3\sigma}+\frac{\epsilon}{\sqrt{\sigma}}\right)}\right\}.
\end{equation}
Now using (\ref{7}) we obtain
$\Omega_{mcg}=\frac{\rho_{mcg}}{\rho}=\frac{u}{u+v}$. So from
equation (\ref{22}) it is obvious that,
\begin{equation}\label{23}
q^{(DGP)}=-1+\frac{3}{2}\left\{\frac{12r_c^2\left(1+\frac{u
w_{mcg}^{(DGP)}}{u+v}\right)}{\sqrt{4r_c^2+3\sigma}\left(\sqrt{4r_c^2+3\sigma}+\frac{\epsilon}{\sqrt{\sigma}}\right)}\right\}
\end{equation}
The solution corresponding to the stable critical point, such that
$\left(u,v\right)\rightarrow(u_{c},v_{c})$. Hence using (\ref{23})
we get
\begin{equation}\label{24}
q_c^{(DGP)}=-1+\frac{3}{2}X_{(DGP)}~~,where~~
X_{(DGP)}=\frac{12r_c^2\left(1+\frac{w_{mcg}^{(DGP)}u_{c}}{u_{c}+v_{c}}\right)}{\left\{\sqrt{4r_c^2+3\sigma}\left(\sqrt{4r_c^2+3\sigma}+\frac{\epsilon}{\sqrt{\sigma}}\right)\right\}}.
\end{equation}

Moreover the Hubble parameter can be obtained as
\begin{equation}\label{25}
H=\frac{2}{3X_{(DGP)}t},
\end{equation}
where we have ignored the integration constant. Integration of
(\ref{25}) yields
\begin{equation}\label{26}
a(t)=a_0t^{\frac{2}{3X_{(DGP)}}},
\end{equation}
which gives a power law form of the expansion. In order to have
accelerating universe $\frac{2}{3X_{(DGP)}}>1~
i.e.,~0<X_{(DGP)}<\frac{2}{3}$. Using this range of $X_{(DGP)}$ in
the equation $q_{c}^{(DGP)}=-1+\frac{3}{2}X_{(DGP)}$. We get the
range of $q_{c}^{(DGP)}$ as $-1<q_{c}^{(DGP)}<0$. Therefore the
deceleration parameter is negative and hence the result is
consistent with the fact that the universe is undergoing an
accelerated expansion.\\

Sahni et al (2003) introduced a pair of cosmological diagnostic
pair $\{r,s\}$ which is known as as statefinder parameters. The
two parameters are dimensionless and are geometrical since they
are derived from the cosmic scale factor alone. Also this pair
generalizes the well-known geometrical parameters like the Hubble
parameter and the deceleration parameter. The statefinder
parameters are given by
\begin{equation}\label{27}
r\equiv\frac{\stackrel{...}a}{aH^3},\ \
s\equiv\frac{r-1}{3(q-1/2)}.
\end{equation}
In the DGP brane model, we have the following expressions of $r$
and $s$ as
\begin{equation}\label{28}
r_{(DGP)}=\left(1-\frac{3X_{(DGP)}}{2}\right)\left(1-3X_{(DGP)}\right).
\end{equation}
and
\begin{equation}\label{29}
s_{(DGP)}=X_{(DGP)}.
\end{equation}

\section{Dynamical system analysis for RS II brane world}

Randall and Sundrum (1999a, 1999b) proposed a bulk-brane model to
explain the higher dimensional theory, popularly known as RS II
brane model. According to this model we live in a four dimensional
world (called 3-brane, a domain wall) which is embedded in a 5D
space time (bulk). All matter fields are confined in the brane
whereas gravity can only propagate in the bulk. The consistency of
this brane model with the expanding universe has given popularity
to this model of late in the field of cosmology.

\subsection{Basic equations in RS II brane model}

In RS II model the effective equations of motion on the 3-brane
embedded in 5D bulk having $Z_{2}$-symmetry are given by
[Maartens, R. 2000, 2004; Randall and Sundrum, 1999b; Shiromizu et
al, 2000; Maeda et al, 2000; Sasaki et al, 2000]
\begin{equation}\label{rsbasic1}
^{(4)}G_{\mu\nu}=-\Lambda_{4}q_{\mu\nu}+\kappa^{2}_{4}\tau_{\mu\nu}+\kappa^{4}_{5}\Pi_{\mu\nu}-E_{\mu\nu}
\end{equation}
where

\begin{equation}\label{rsbasic2}
\kappa^{2}_{4}=\frac{1}{6}~\lambda\kappa^{4}_{5}~,
\end{equation}
\begin{equation}\label{rsbasic3}
\Lambda_{4}=\frac{1}{2}~\kappa^{2}_{5}\left(\Lambda_{5}+\frac{1}{6}~\kappa^{2}_{5}\lambda^{2}\right)
\end{equation}
and
\begin{equation}\label{rsbasic4}
\Pi_{\mu\nu}=-\frac{1}{4}~\tau_{\mu\alpha}\tau^{\alpha}_{\nu}+\frac{1}{12}~\tau\tau_{\mu\nu}+\frac{1}{8}~
q_{\mu\nu}\tau_{\alpha\beta}\tau^{\alpha\beta}-\frac{1}{24}~q_{\mu\nu}\tau^{2}
\end{equation}
and $E_{\mu\nu}$ is the electric part of the 5D Weyl tensor. Here
$\kappa_{5},~\Lambda_{5},~\tau_{\mu\nu}$ and $\Lambda_{4}$ are
respectively the 5D gravitational coupling constant, 5D
cosmological constant, the brane tension (vacuum energy), brane
energy-momentum tensor and effective 4D cosmological constant.
The explicit form of the above modified Einstein equations in flat universe are

\begin{equation}\label{rsbasic5}
3H^{2}=\Lambda_{4}+\kappa^{2}_{4}\rho+\frac{\kappa^{2}_{4}}{2\lambda}~\rho^{2}+\frac{6}{\lambda
\kappa^{2}_{4}}\cal{U}
\end{equation}
and
\begin{equation}\label{rsbasic6}
2\dot{H}+3H^{2}=\Lambda_{4}-\kappa^{2}_{4}p-\frac{\kappa^{2}_{4}}{2\lambda}~\rho
p-\frac{\kappa^{2}_{4}}{2\lambda}~\rho^{2}-\frac{2}{\lambda
\kappa^{2}_{4}}\cal{U}
\end{equation}
The dark radiation $\cal{U}$ obeys

\begin{equation}\label{rsbasic7}
\dot{\cal U}+4H{\cal U}=0
\end{equation}
where $\rho=\rho_{mcg}+\rho_{m}$ and $p=p_{mcg}+p_{m}$ are the
total energy density and pressure respectively. The continuity
equations for dark energy and dark matter are given in (\ref{4})
and (\ref{5}). Now we shall study the dynamical system assuming
$\Lambda_{4}={\cal U}=0$ (in absence of cosmological constant and
dark radiation).

\subsection{Dynamical system analysis}

Now to analyze the dynamical system we define three variables $x$,
$u$ and $v$ as given in equation (\ref{7}). Now making use of
(\ref{4}), (\ref{5}), (\ref{7}), (\ref{rsbasic5}) and
(\ref{rsbasic6}) we get,
\begin{equation}\label{rsbasic8}
\frac{du}{dx}=-3b\left(u+v\right)-3u\left(1+w_{mcg}^{(RSII)}\right)+3\kappa_{4}^{2}u\left\{u\left(1+w_{mcg}^{(RSII)}\right)+v\right\}-\frac{3u}{u+v}\left\{1-\kappa_{4}^{2}(u+v)\right\}\left\{u\left(2+w_{mcg}^{(RSII)}\right)+2v\right\}
\end{equation}
and
\begin{equation}\label{rsbasic9}
\frac{dv}{dx}=3b\left(u+v\right)-3v+3v\left[\kappa_{4}^{2}\left\{u\left(1+w_{mcg}^{(RSII)}\right)+v\right\}-\frac{3}{u+v}\left\{1-\kappa_{4}^{2}(u+v)\right\}\left\{u\left(2+w_{mcg}^{(RSII)}\right)+2v\right\}\right]
\end{equation}
where
$$w_{mcg}^{(RSII)}=A-\left[\frac{B\kappa_{4}^{2}\left(u+v\right)^{2}}{2u\lambda\left\{1-\kappa_{4}^{2}\left(u+v\right)\right\}}\right]^{\alpha+1}$$

\subsubsection{Critical point}

Similar to DGP brane model, the critical points of the above
system are obtained by putting $\frac{du}{dx}=0=\frac{dv}{dx}$.
Now this system of equation does not yield an explicit solution.
So we have to put some special values to the constants which will
yield non-zero positive solution of the system of equations.
Considering $A=0, B=1, b=0.5, \lambda=1$, $\alpha=1$ and
$\kappa=1$ we obtain the following critical points to the system.
\begin{equation}\label{rsbasic9A}
u_{1c}=0.491891,~~~~v_{1c}=0.190983
\end{equation}
\begin{equation}\label{rsbasic9B}
u_{2c}=1.19631,~~~~v_{2c}=0.836396
\end{equation}
\begin{equation}\label{rsbasic9C}
u_{3c}=2.22826,~~~~v_{3c}=10.9091
\end{equation}

\subsubsection{Stability around critical point}

Now if we write $\tilde{f}=\frac{du}{dx}$ and
$\tilde{g}=\frac{dv}{dx}$, then we can obtain the following
expressions
\begin{equation}\label{rsbasic10}
\delta\left(\frac{du}{dx}\right)=\left[\partial_{u}
\tilde{f}\right]_{c}\delta u+\left[\partial_{v}
\tilde{f}\right]_{c}\delta v
\end{equation}
and
\begin{equation}\label{rsbasic11}
\delta\left(\frac{dv}{dx}\right)=\left[\partial_{u}
\tilde{g}\right]_{c}\delta u+\left[\partial_{v}
\tilde{g}\right]_{c}\delta v
\end{equation}
where

$$\partial_{u} \tilde{f} = \frac{3}{4u^{2}\left(u+v\right)^{2}\left\{(u+v)k^{2}-1\right\}^{3}\lambda^{2}}\left[B(u+v)^{4}k^{4}\left\{-6u^{2}-2uv+v^{2}+(u+v)\left(10u^{2}-v^{2}\right)k^{2}-4u^{2}(u+v)^{2}k^{4}\right\}\right.$$
\begin{equation}\label{26a}\left.+4u^{2}\left\{\left(u+v\right)k^{2}-1\right\}^{3}\left\{-3u^{2}-2Au^{2}-bu^{2}-6uv-4Auv-2buv-3v^{2}-Av^{2}-bv^{2}+\left(u+v\right)^{2}\left(\left(6+4A\right)u+3v\right)k^{2}\right\}\lambda^{2}\right]\end{equation}

$$\partial_{v} \tilde{f} = \frac{3}{4u\left(u+v\right)^{2}\left\{(u+v)k^{2}-1\right\}^{3}\lambda^{2}}\left[B(u+v)^{4}k^{4}\left\{-7u-4v+\left(u+v\right)\left(11u+2v\right)k^{2}-4u\left(u+v\right)^{2}k^{4}\right\}\right.$$
\begin{equation}\label{26b}\left.+4u\left\{\left(u+v\right)k^{2}-1\right\}^{3}\left\{Au^{2}+\left(u+v\right)^{2}\left(-b+3uk^{2}\right)\right\}\lambda^{2}\right]\end{equation}

\begin{equation}\label{26c}
\partial_{u} \tilde{g} = 3\left[b+v\left\{\left(7+4A\right)k^{2}-\frac{3Av}{\left(u+v\right)^{2}}\right\}+\frac{Bv\left(u+v\right)^{2}k^{4}\left\{-6u+3v+\left(12u-7v\right)\left(u+v\right)k^{2}+4\left(v-u\right)\left(u+v\right)^{2}k^{4}\right\}}{4u^{2}\left\{\left(u+v\right)k^{2}-1\right\}^{3}\lambda^{2}}\right]
\end{equation}

$$\partial_{v} \tilde{g} = \frac{3}{4u\left(u+v\right)^{2}\left\{(u+v)k^{2}-1\right\}^{3}\lambda^{2}}\left[B\left(u+v\right)^{4}k^{4}\left\{-3\left(u+4v\right)+\left(u+v\right)\left(7u+26v\right)k^{2}-4\left(u+v\right)^{2}\left(u+3v\right)k^{4}\right\}\right.$$
\begin{equation}\label{26d}\left.+4u\left\{\left(u+v\right)k^{2}-1\right\}^{3}\left\{-7u^{2}-3Au^{2}+bu^{2}-14uv+2buv-7v^{2}+bv^{2}+\left(u+v\right)^{2}\left\{\left(7+4A\right)u+14v\right\}k^{2}\right\}\lambda^{2}\right]\end{equation}

\begin{figure}

\includegraphics[height=2in]{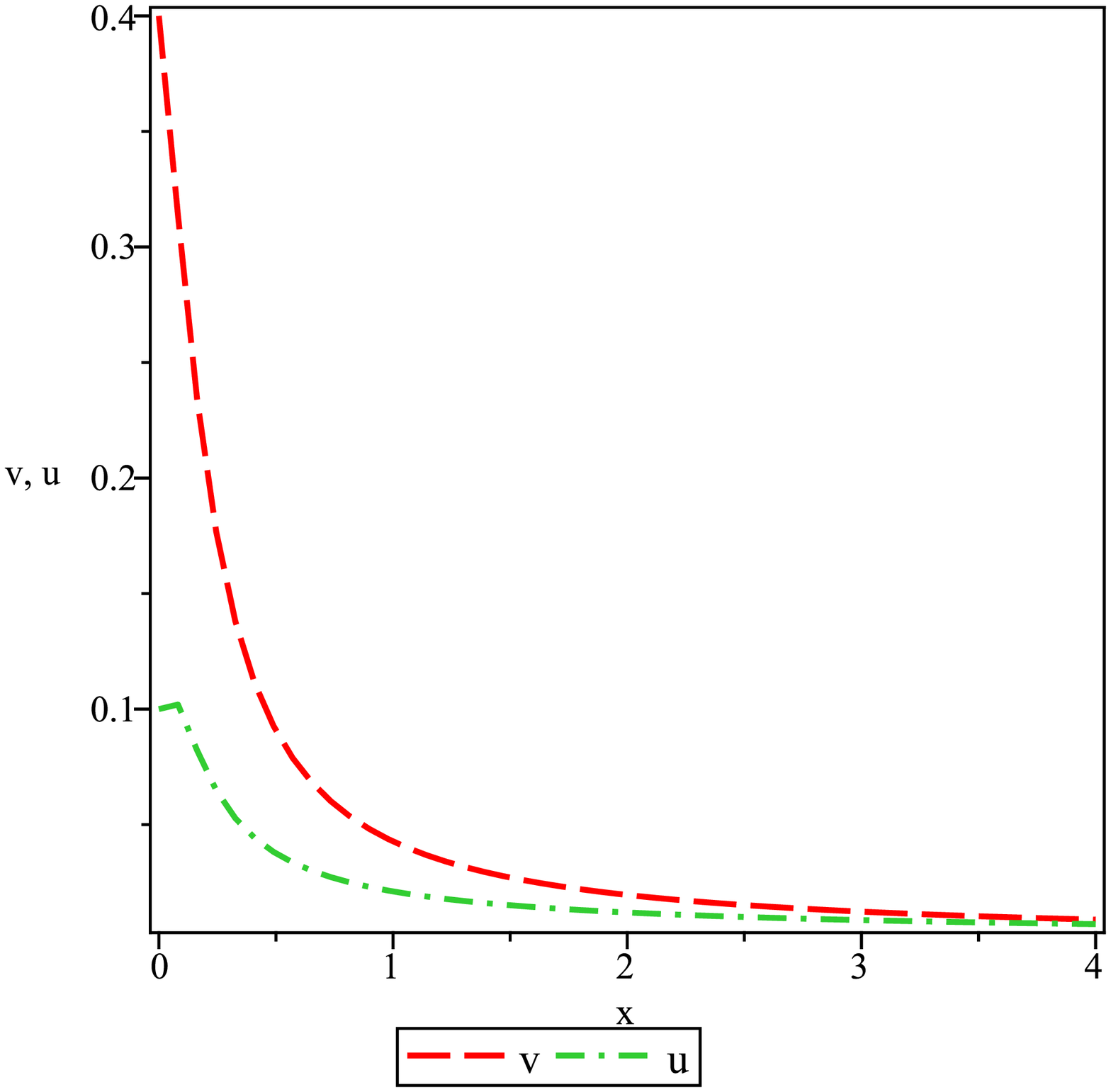}~~~~~~~~\includegraphics[height=2in]{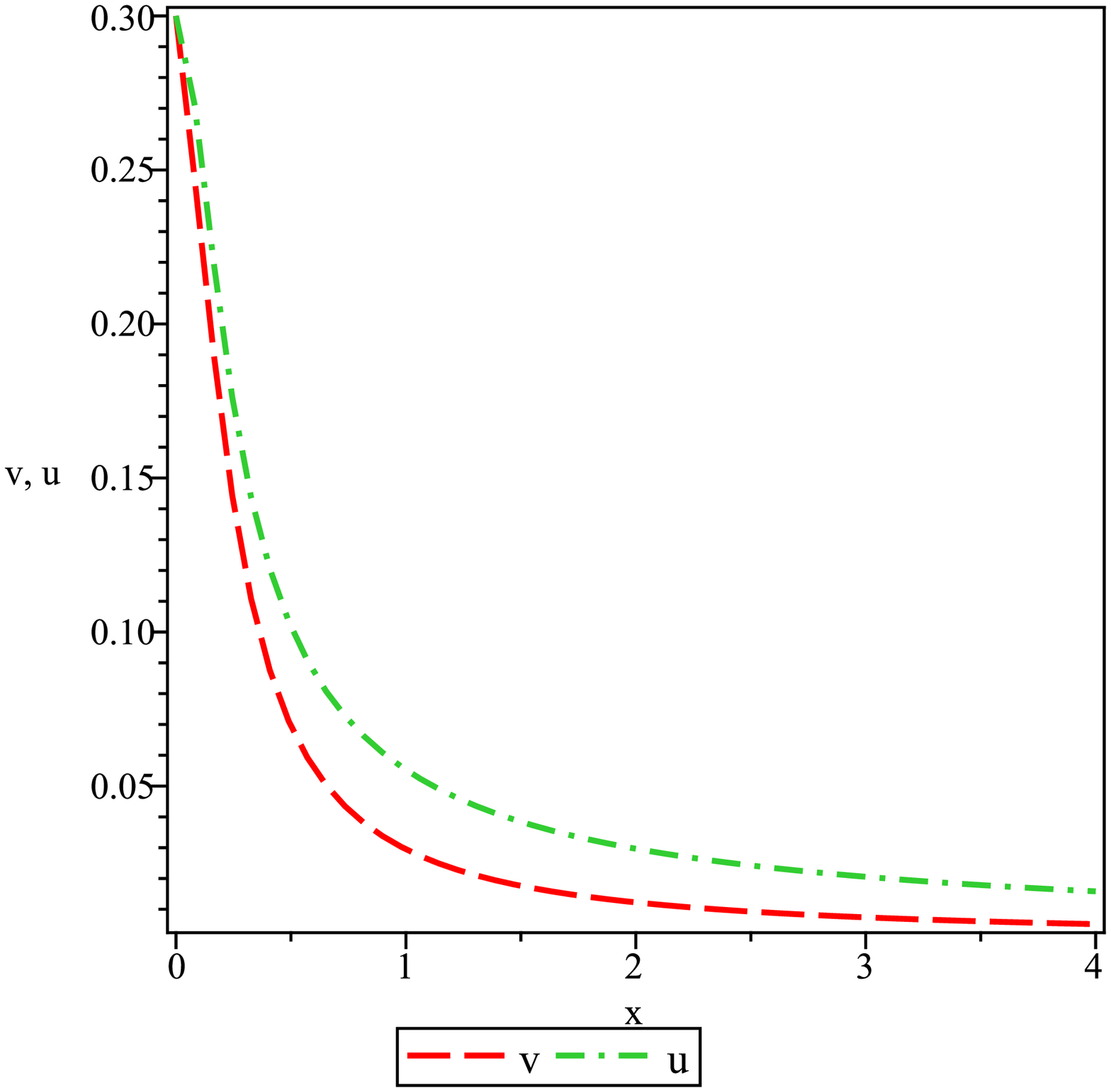}~~~~~~~~\includegraphics[height=2in]{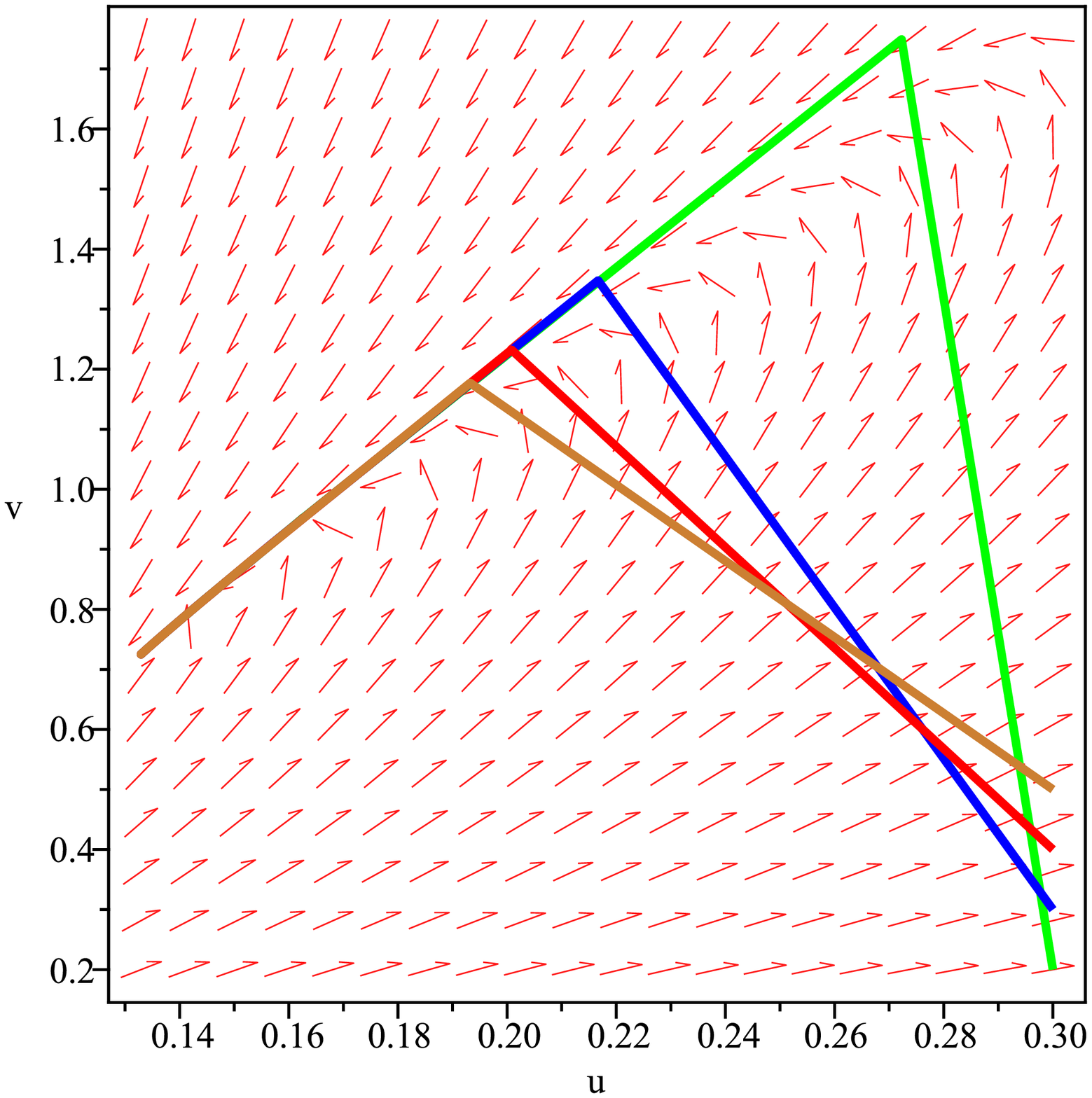}\\
\vspace{1mm}
~~~~~~~~~~~~~~~~Fig. 8~~~~~~~~~~~~~~~~~~~~~~~~~~~~~~~~~~~~~~Fig. 9~~~~~~~~~~~~~~~~~~~~~~~~~~~~~~~~~~~~~~~~~~Fig. 10~~~\\
\vspace{2mm}

Fig 8 : The dimensionless density parameters are plotted against
e-folding time. The initial condition is $v(0)=0.4, u(0)=0.1$.
Other parameters are fixed at $\alpha=1, b=0.5, A=0, B=1,
\lambda=1$, and $\kappa=1$.\\Fig 9 :
 The dimensionless density parameters are
plotted against e-folding time.The initial condition is $v(0)=0.3,
u(0)=0.3$. The other parameters are fixed at $\alpha=1, b=0.5,
A=0, B=1, \lambda=1$ and $\kappa=1$.\\Fig 10 :  The phase diagram
of the parameters depicting an attractor solution. The initial
conditions chosen are $v(0)=0.2, u(0)=0.3$ (green); $v(0)=0.3,
u(0)=0.3$ (blue); $v(0)=0.4, u(0)=0.3$ (red); $v(0)=0.5, u(0)=0.3$
(brown). Other parameters are fixed at $\alpha=1, b=0.01, A=0.3,
B=1, \lambda=1$ and $\kappa=5$.\\\\\\\\

\includegraphics[height=1.5in]{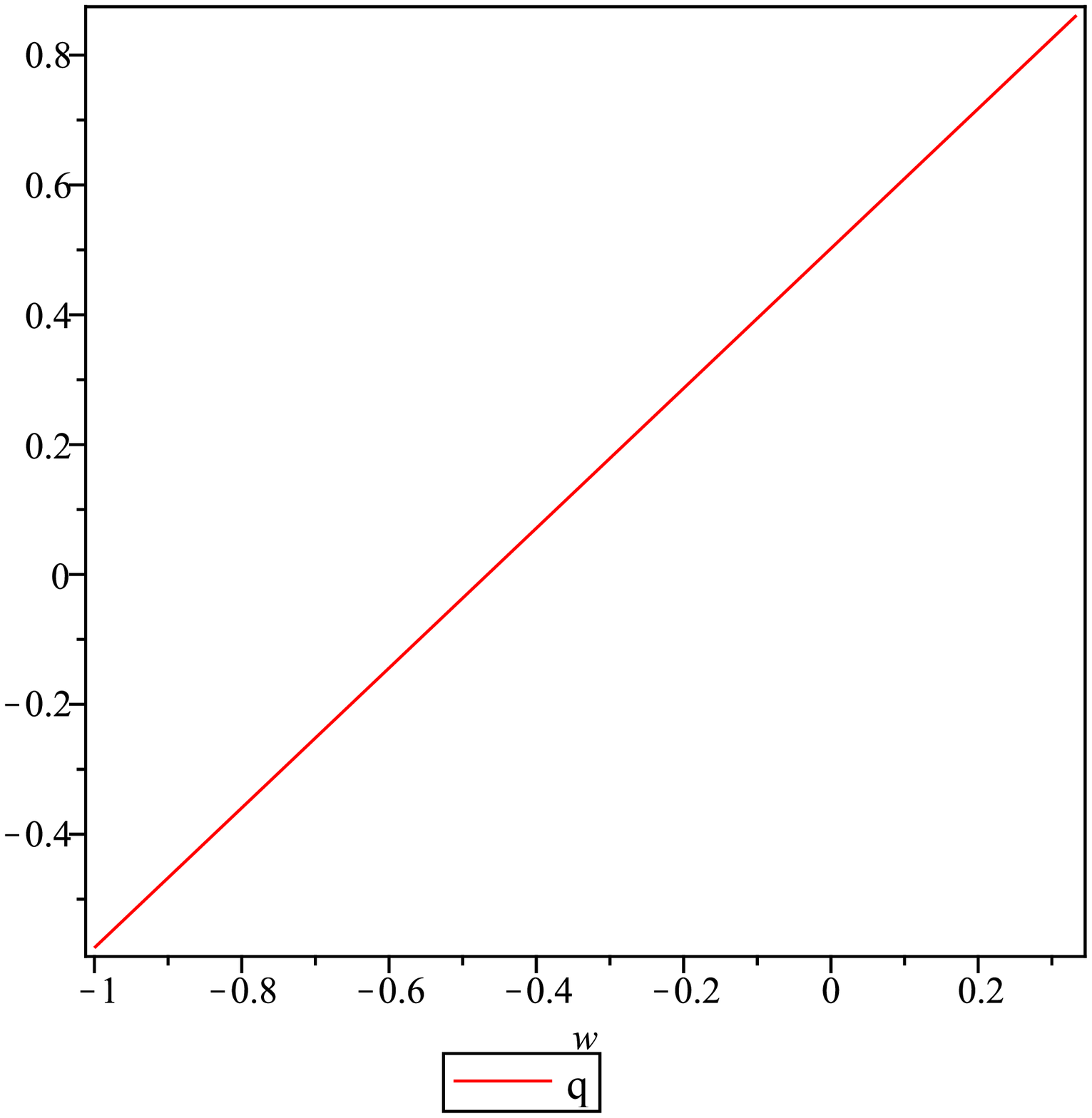}~~~~~\includegraphics[height=1.5in]{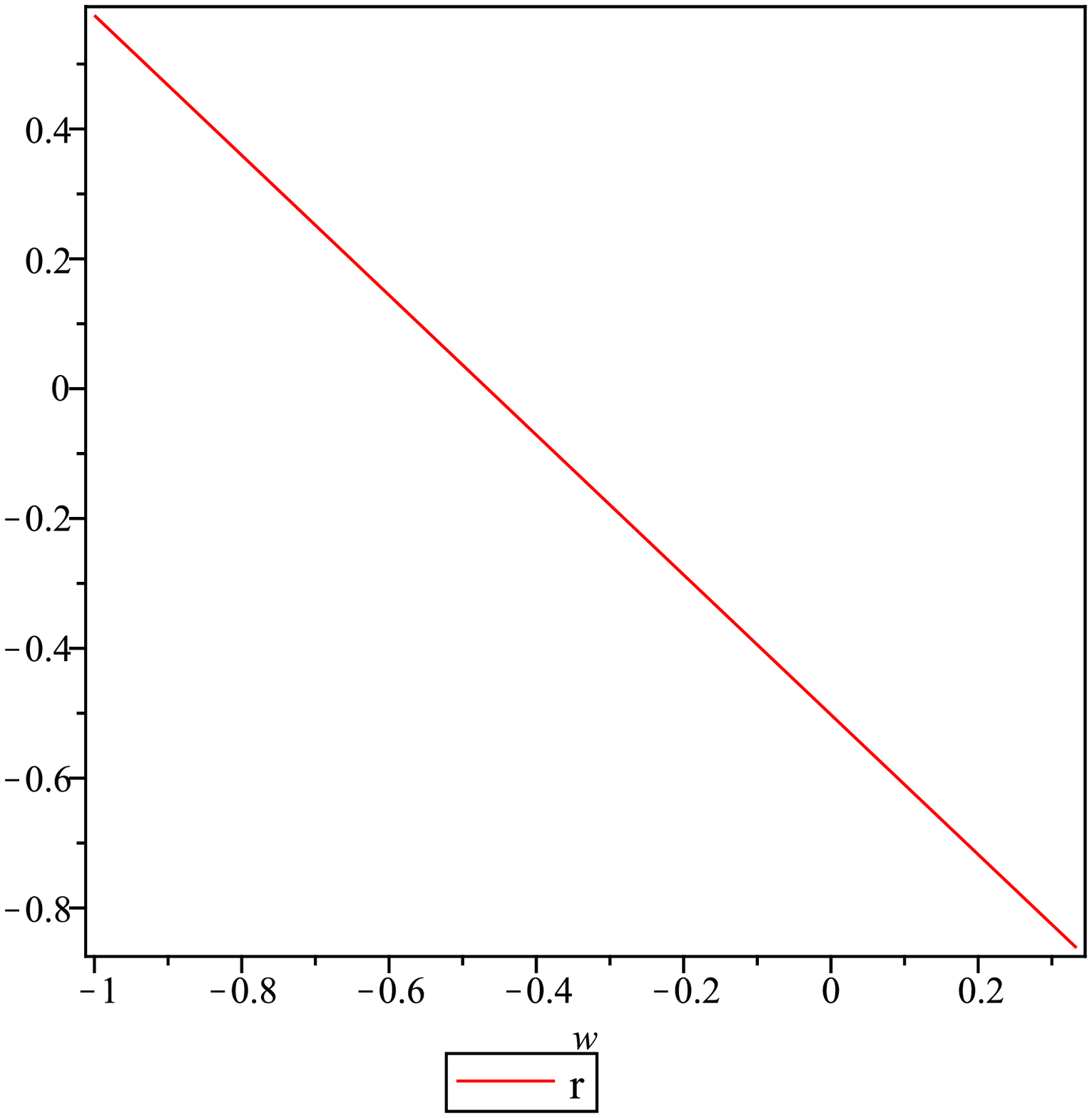}
~~~~~\includegraphics[height=1.5in]{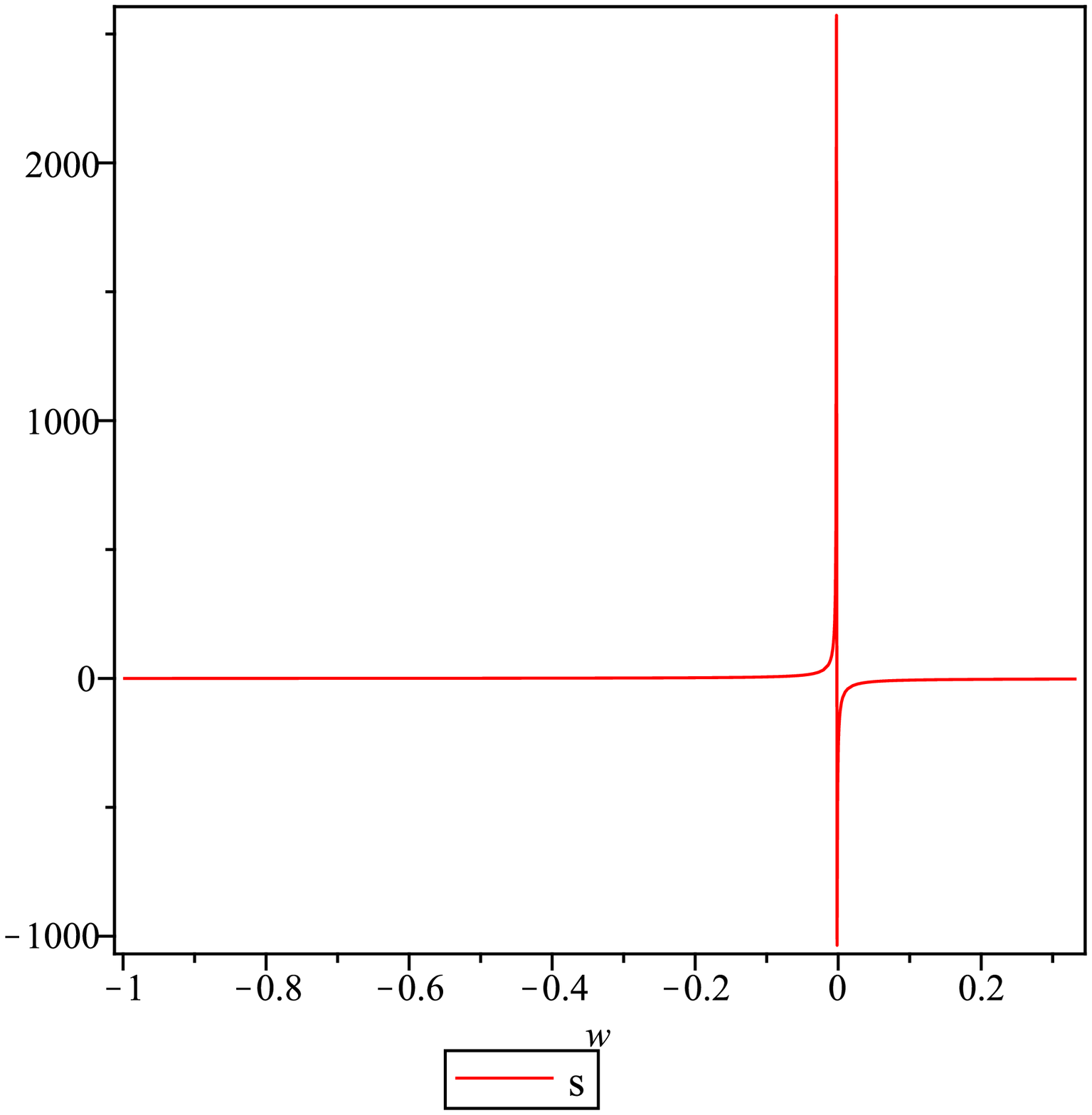}~~~~~\includegraphics[height=1.5in]{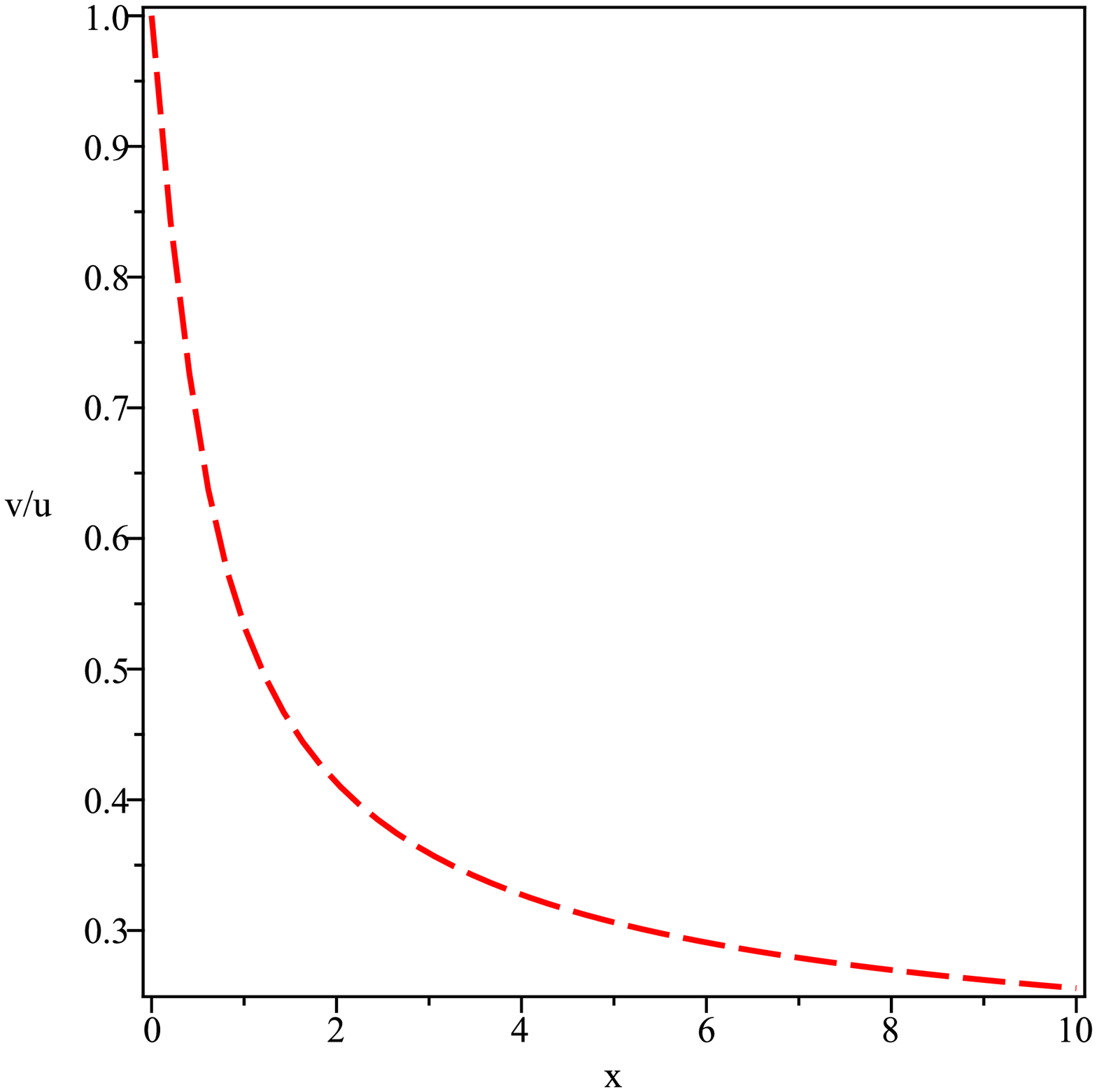}~\\
\vspace{1mm}
~~~~~~~~~Fig. 11~~~~~~~~~~~~~~~~~~~~~~~~~Fig. 12~~~~~~~~~~~~~~~~~~~~~~~~~~~Fig. 13~~~~~~~~~~~~~~~~~~~~~~~~~~Fig. 14\\
\vspace{2mm}

Fig 11 :   The deceleration parameter is plotted against the state
parameter. $\epsilon=0.003$ is considered.
\\Fig 12 :  The statefinder parameter $r$ is plotted
against the state parameter. $\epsilon= 0.003$ is considered.
\\ Fig 13 : The statefinder parameter $s$ is plotted
against the state parameter. $\epsilon=0.003$ is considered.
\\Fig 14 : The ratio of density parameters is shown
against e-folding time. The initial conditions chosen are
$v(0)=0.3, u(0)=0.3$. The other parameters are fixed at $\alpha=1,
A=0, b=0.5, B=1, \lambda=1$ and $\kappa=1$.
\end{figure}

\subsubsection{Nature of cosmological parameters}

In this RSII model, the deceleration parameter $q$ can be obtained
as
\begin{equation}\label{rsbasic12}
q^{(RSII)}=-1-\frac{3}{2}\frac{\left\{\frac{\rho}{2\lambda}\left(w_{mcg}^{(RSII)}\frac{\rho_{mcg}}{\rho}-2\right)-\left(1+w_{mcg}^{(RSII)}\frac{\rho_{mcg}}{\rho}\right)\right\}}{\left(1+\frac{\rho}{2\lambda}\right)}
\end{equation}
which can be written in terms of dimensionless density parameter
$\Omega_{mcg}=\frac{\rho_{mcg}}{\rho}$ as in the following
\begin{equation}\label{rsbasic13}
q^{(RSII)}=-1+\frac{3}{2}\frac{\left\{\left(\frac{-\rho}{2\lambda}+1\right)w_{mcg}^{(RSII)}\Omega_{mcg}+\left(1+\frac{\rho}{\lambda}\right)\right\}}{\left(1+\frac{\rho}{2\lambda}\right)}
\end{equation}
Now since  $\Omega_{mcg}=\frac{\rho_{mcg}}{\rho}=\frac{u}{u+v}$
and assuming $\frac{\rho}{\lambda}=\epsilon_{(RSII)}$ we get,
\begin{equation}\label{rsbasic14}
q^{(RSII)}=-1+\frac{3}{2}\frac{\left\{\left(1-\frac{\epsilon_{(RSII)}}{2}\right)w_{mcg}^{(RSII)}\frac{u}{u+v}
+\left(1+\epsilon_{(RSII)}\right)\right\}}{\left(1+\frac{\epsilon_{(RSII)}}{2}\right)}
\end{equation}
Considering only the first stable critical point, such that
$(u,v)\rightarrow(u_{1c},v_{1c})$, using (\ref{rsbasic14}) we get,
\begin{equation}\label{rsbasic15}
q_c^{(RSII)}=-1+\frac{3}{2}X_{(RSII)},~~~ where ~~~~
X_{(RSII)}=\frac{\left\{\left(1-\frac{\epsilon_{(RSII)}}{2}\right)w_{mcg}^{(RSII)}\frac{u_{1c}}{u_{1c}+v_{1c}}
+\left(1+\epsilon_{(RSII)}\right)\right\}}{\left(1+\frac{\epsilon_{(RSII)}}{2}\right)}
\end{equation}
If
$\epsilon_{(RSII)}=-\frac{2\left[\left(1+w_{mcg}^{(RSII)}\right)u_{1c}+v_{1c}\right]}{\left(2-w_{mcg}^{(RSII)}\right)u_{1c}+2v_{1c}},
~~X_{(RSII)}=0, ~~~$ we have $q=-1$, which confirms the
accelerated expansion  of the universe. When
$\epsilon_{(RSII)}=-2$ we have $q=-\infty$. Therefore we have
super accelerated expansion of the universe.\\

In this scenario, the Hubble parameter can be obtained as,
\begin{equation}\label{rsbasic16}
H=\frac{2}{3X_{(RSII)}t}
\end{equation}
where the integration constant has been ignored. Integration of
(\ref{rsbasic16}) yields
\begin{equation}\label{rsbasic17}
a(t)=a_0t^{\frac{2}{3X_{(RSII)}}}
\end{equation}
which gives the power law form of expansion of the universe. Like
DGP brane here also in order to have an accelerated expansion of
universe in RSII brane we must have $0<X_{(RSII)}<\frac{2}{3}$.
Using this range of $X_{(RSII)}$ in the equation
$q_{c}^{RSII}=-1+\frac{3}{2}X_{(RSII)}$, we get the range of
$q_{c}^{(RSII)}$ as $-1<q_{c}^{(RSII)}<0$. This is again
consistent with an accelerated expansion of the universe as it was
in DGP model. Like before here also we calculate the statefinder
parameters $\{r,s\}$ in order to get relevant information of DE
and DM in the context of background geometry only without
depending on the theory of gravity. The expressions of the
statefinder pair (eq. (\ref{27})) in the RSII model can be
obtained in the form
\begin{equation}\label{rsbasic19}
r_{(RSII)}=\left(1-\frac{3X_{(RSII)}}{2}\right)\left(1-3X_{(RSII)}\right).
\end{equation}
and
\begin{equation}\label{rsbasic20}
s_{(RSII)}=X_{(RSII)}
\end{equation}

\section{Graphical Representation of the Phase plane Analysis}

Phase diagrams are drawn to determine the type of critical point
in DGP and RSII models. We discuss the results obtained in the two
models separately.

$$DGP~~BRANE~~MODEL$$

The dimensionless density parameters $v$ and $u$ are drawn in
figures 1 and 2. From the figures we see that $v$ decreases, and
$u$ increases during evolution of the universe. This shows that
the density of DM decreases while the density of DE increases as
the universe evolves. So this result is consistent with the well
known idea of an energy dominated universe. The phase space
diagram (figure 3) shows the attractor solution. The eigen values
are calculated at the critical point and found to be $(1.97625,~
-1.09495)$. Hence the critical point is a saddle
point.\\

Figure 4 shows the variations of the deceleration parameter, $q$
against $w_{mcg}$. From figure 4, it is evident that there is a
gradual decrease in the deceleration parameter $q$, and finally in
the late universe it attains negative values, which suggests that
there should be an acceleration in the late universe. This result
is in accordance with the various observational data of Ia
supernovae and CMB data which suggests that the universe is
undergoing an accelerated expansion of late. Figure 5 and 6 shows
the variations of the statefinder parameters $r$ and $s$ against
$w_{mcg}$. From these figures it can be seen that $r$ tends
towards $1$ and $s$ tends towards $0$. Therefore it is evident
that these results tends towards the $\Lambda CDM$ model. Figure 7
shows the variation of the ratio of the density parameter $v$ and
$u$ with time. From the figure it is evident that the ratio of the
above parameters decreases with time. So it can be concluded that
there is a relative decrease in matter density with respect to the
energy density. This is again consistent with the notion of an
energy dominated universe.

$$RSII~~BRANE~~MODEL$$

The dimensionless density parameters $v$ and $u$ in case of RSII
brane model are drawn in figures 8 and 9. From the figures it is
evident that at early stages of evolution of universe density of
DM decreases steeply relative to the density of DE such that
finally an energy dominated scenario is witnessed in the late
universe. The phase space diagram (figure 10) shows the attractor
solution for the model. The eigen values are calculated at all the
three critical points. At the first critical point the eigen
values are found to be $(402.877,~ 204.835)$. Hence the first
critical point is an unstable node. At the second critical point
the eigen values are found to be $(1214.75,~ 565.097)$. Hence the
second critical point is also an unstable node. At the third
critical point the eigen values calculated are $(-25.3905 +
2257.49i,~ -25.3905 -2257.49i )$. Hence the third critical point is a stable focus.\\

Figure 11 shows the behaviour of deceleration parameter $q$ as it
is plotted against the state parameter $w_{mcg}$. The figure shows
that there is a gradual decrease in the deceleration parameter
with time. Finally in the late universe the deceleration parameter
$q$ becomes negative, which confirms the fact that the universe is
undergoing an accelerated expansion. In Figures 12 and 13, the
statefinder parameters $r$ and $s$ are plotted against $w_{mcg}$.
Finally in Figure 14 the ratio of the density parameters $v$ and
$u$ is plotted against time. From the figure it is found that just
like DGP brane here also there is a relative decrease in matter
density with respect to the energy density. This is consistent
with the notion of an energy dominated universe. Henceforth we
find a solution for the cosmic coincidence problem.\\\\

\section{Study of Future Singularities}

It is a well known fact that any energy dominated model of the
universe is destined to result in a future singularity. The study
of dynamics of an accelerating universe in the presence of DE and
DM is in fact incomplete without studying these singularities,
which are the ultimate fate of the universe. MCG, being a very
generalized form of DE can step into the phantom era with the EoS
$(w<-1)$ for definite values of  the constants $A$ and $B$. It is
known that the universe dominated by phantom energy ends with a
future singularity known as Big Rip (Caldwell, R. R., 2003), due
to the violation of dominant energy condition (DEC). But this is
not the only type of singularity possible. Nojiri et al (2005)
studied the various types of singularities that can result from an
phantom energy dominated universe. These possible singularities
are characterized by the growth of energy and curvature at the
time of occurrence of the singularity. It is found that near the
singularity quantum effects becomes very dominant which may
alleviate or even prevent these singularities. So it is extremely
necessary to study these singularities and classify them
accordingly so that we can search for methods to eliminate them.
The appearance of all four types of future singularities in
coupled fluid dark energy, $F(R)$ theory, modified Gauss-Bonnet
gravity and modified $F(R)$ Horava-Lifshitz gravity was
demonstrated in (Nojiri, S. and Odintsov, S. D., 2011). The
universal procedure for resolving such singularities that may lead
to bad phenomenological consequences was proposed. Nojiri, S. and
Odintsov, S. D., 2011 classified and studied four types of such
singularities in detail. We analyze two possible future singularities
in case of MCG in brane world scenarios.\\
Type $I$ (``Big Rip") :
For $t\rightarrow t_{s}~,~a\rightarrow \infty~,~\rho\rightarrow \infty~and~|p| \rightarrow \infty$\\
Type $II$ (``Sudden") : For $t\rightarrow t_{s}~,~a\rightarrow a_{s}~,~\rho\rightarrow \rho_{s}~and~|p| \rightarrow \infty$\\

Here we have considered MCG as the DE in DGP and RS2 brane models.
1)  Phantom era is realized for MCG when $\rho<
[\frac{B}{A+1}]^\frac{1}{\alpha+1}$. From the MCG equation of
state i.e $p=A\rho-\frac{B}{\rho^\alpha}$ we see that when
$\rho\rightarrow\infty$ , $|p|\rightarrow\infty$  when
$a\rightarrow\infty$ and $t\rightarrow t_{s}$. Therefore it is
seen that Type I singularity is quite obvious for MCG. This is the
well known Big Rip singularity.

2)  Suppose if $\rho\rightarrow\rho_{s}$ and $\rho_{s}\sim0$, then
$|p|\rightarrow-\infty$ for $t\rightarrow t_{s}$ and $a\rightarrow
a_{s}$ Hence we get the type II singularity. This is called a
sudden singularity.

The stability of the critical points analyzed in the last section
will be occurred much before than the singularities take place, so
any direct impact of the singularities of the dynamical system
will not be observed. The big rip and sudden singularity may be
formed before the critical points to be formed.

\section{Concluding Remarks}

In this work, we have considered a combination of Modified
Chaplygin gas in DGP and RSII brane gravity models. Our basic idea
was to study the background dynamics of MCG in detail when it is
incorporated in brane gravity. Dynamical system analysis had been
carried out, critical points were found and the stability of the
system around those critical points was tested for both DGP and
RSII brane models. Graphical analysis was done to get an explicit
picture of the outcome of the work. In order to find a solution
for the cosmic coincidence problem, a suitable interaction between
DE and DM was considered. The dynamical system of equations
characterizing the system was formed and a stable scaling solution
was obtained. The numerical values of the density parameters for
DE and DM almost coincided with each other (as seen in graphs 8
and 9). Hence this work can be considered to be a significant one
as far as solution of cosmic coincidence problem is concerned.
From the above analysis we conclude that the combination of MCG in
brane gravity makes a perfect model for the expanding universe
undergoing a late acceleration. \\

{\bf Achknowledgement:}\\\\ RB thanks to State Govt. of West
Bengal, India for awarding SRF.

\end{document}